\documentclass[12pt]{article}
\usepackage{amssymb}
\usepackage{graphics}
\usepackage{epsfig}
\usepackage{a4wide}

\begin{document}
\newcommand{\pphww}{$pp \to H^0 W^+ W^-+X~$ }
\newcommand{\pphwwg}{$pp \to H^0 W^+ W^-g+X~$ }
\newcommand{\qqhww}{$q\bar q \to H^0W^+W^-~$}
\newcommand{\qqhwwg}{$q\bar q \to H^0W^+W^-g~$}
\newcommand{\gghww}{$gg \to H^0W^+W^-~$}
\newcommand{\qghwwq}{$q(\bar q)g \to H^0W^+W^-q(\bar q)~$}
\newcommand{\ppuuhww}{$pp \to u\bar u \to H^0W^+W^-+X~$}
\newcommand{\ppqqhww}{$pp \to q\bar q \to H^0W^+W^-+X~$}
\newcommand{\ppgghww}{$pp \to gg \to H^0W^+W^-+X~$}

\title{ QCD corrections to associated Higgs boson production
with a W boson pair at the LHC }
\author{ Song Mao, Ma Wen-Gan, Zhang Ren-You, Guo Lei, \\
         Wang Shao-Ming, and Han Liang \\
{\small Department of Modern Physics, University of Science and Technology}\\
{\small of China (USTC), Hefei, Anhui 230026, P.R.China}  }

\date{}
\maketitle \vskip 15mm
\begin{abstract}
The Higgs boson production in association with a pair of W-bosons at
the Large Hadron Collider(LHC) can be used to probe the coupling
between Higgs boson and vector gauge bosons and discover the
signature of new physics. We describe the impact of the complete QCD
NLO radiative corrections and the gluon-gluon fusion subprocesss to
the cross section of this process at the LHC, and investigate the
dependence of the leading order(LO) and the QCD corrected cross
sections on the fctorization/renormalization energy scale and Higgs
boson mass. We present the LO and QCD corrected distributions of the
invariant mass of W-boson pair and the transverse momenta of final W
and Higgs boson. We find that the QCD NLO corrections and the
contribution from gluon-gluon fusion subprocess significantly modify
the LO distributions, and the scale dependence of the QCD corrected
cross section is badly underestimated by the LO results. Our
numerical results show that the K-factor of the QCD correction
varies from $1.48$ to $1.64$ when $m_H$ goes up from $100~GeV$ to
$160~GeV$. We find also the QCD correction from \gghww subprocess at
the LHC is significant, and should be considered in precise
experiment.
\end{abstract}


\vfill \eject

\baselineskip=0.32in

\renewcommand{\theequation}{\arabic{section}.\arabic{equation}}
\renewcommand{\thesection}{\Roman{section}.}
\newcommand{\nb}{\nonumber}

\newcommand{\Dir}{\kern -6.4pt\Big{/}}
\newcommand{\Dirin}{\kern -10.4pt\Big{/}\kern 4.4pt}
\newcommand{\DDir}{\kern -7.6pt\Big{/}}
\newcommand{\DGir}{\kern -6.0pt\Big{/}}

\makeatletter      
\@addtoreset{equation}{section}
\makeatother       

\section{Introduction}
\par
The Higgs mechanism plays a crucial role in the standard model(SM).
The existence of the Higgs boson makes the breaking of the
electroweak(EW) symmetry and generates the masses for the
fundamental particles \cite{sm,higgs}. Therefore, to study the Higgs
mechanism is one of the main goals of the LHC. The LEP experimental
data from direct search for Higgs boson in association with $Z^0$
boson provide the exclusion of the Higgs boson in the mass range up
to $114.4~ GeV$ at $95\%$ confidence level(CL)\cite{lower mH}. The
current SM fit of all electroweak parameters produced by the LEP
Electroweak Group predicts
$m_H=84^{+34}_{-26}~GeV$\cite{HiggsBounds}, or the one-sided $95\%$
CL limit $m_H < 154~GeV$. Including the LEP direct search results,
this upper limit increases to $m_H \lesssim 185~GeV$\cite{upper mH}.
It is also interesting that recent combined results from the
Tevatron experiments have, for the first time, excluded the
hypothesis of a Higgs boson mass around $170~GeV$\cite{herndon} at
$95\%$ CL. Although the expected sensitivity of Tevatron experiments
is not enough to make a $5\sigma$ discovery of the SM Higgs
boson\cite{rembold}, it is enough to exclude it out up to $m_{H}
\sim 200~GeV$ at $95\%$ CL, or to make a $3\sigma$ observation.
While for the coupling properties, such as the couplings between
Higgs boson and gauge bosons, the precise data provide only little
information about them.

\par
The CERN Large Hadron Collider(LHC) is a machine with the entire
proton-proton colliding energy of $14~TeV$ and a luminosity of 100
$fb^{-1}$ per year. If the Higgs boson really exists, it will be
discovered at the LHC, which can provide a measurement of the Higgs
mass at the per-mille level, and of the Higgs boson coupling at the
$5-20\%$ level. At this machine, the Higgs boson production is
dominated by the gluon-gluon fusion process, described at the
leading order through a heavy-quark loop. The next-to-leading order
cross section for this process is $37.6~pb$, for $m_H=120~GeV$. The
Higgs boson can also be produced by Vector Boson Fusion (VBF) with a
cross section of $4.25~pb$, or by associated production with a
$W^{\pm}$, a $Z^0$, or a $t\bar{t}$ quark pair, with $3.19 \,pb$ for
the three processes and $m_H=120~GeV$ (cross sections calculated at
next-to-leading order)\cite{cscbook}).

\par
After the discovery of Higgs boson, our main task is to probe its
properties, such as spin, CP, and couplings. However, these
measurements require accurate theoretical predictions for both
signal and background. The process \pphww is one of the important
processes in providing the detail information about the coupling
between Higgs-boson and vector gauge bosons. As we will see in this
work, the QCD corrections increase the $H^0W^+W^-$ cross section
significantly, and thus in the quantitative measurement of the
coupling $H^0W^+W^-$ we have to take the QCD corrections into
account.

\par
At the LHC, most of the important processes will involve
multi-particle final states, either through the direct
multi-particle production or the decay of resonances. It is known
that the theoretical predictions beyond the LO for these processes
with more than two final particles are necessary from the data
analysis point of view in order to probe the SM and find new
physics, but the calculations for these processes involving the NLO
corrections are very intricate. In the last few years, the
phenomenological results including the QCD NLO corrections for
tri-boson production processes at the LHC, such as $pp\to
W^+W^-Z^0,~ H^0H^0H^0, ~Z^0Z^0Z^0,$ have been provided
\cite{Lazopoulos:2007ix,Hankele:2007sb,Plehn:2005nk,Binoth:2006ym}.
The QCD NLO corrections to the weak boson fusion processes, like
$pp\to WWjj, WZjj$ \cite{Jager:2006zc,Bozzi:2007ur}, $pp\to Hjj$
with effective gluon-Higgs coupling, \cite{Campbell:2006xx} $gg \to
Hq\bar{q} $\cite{Weber:2006au}, and $pp \to t\bar{t}j $
\cite{Dittmaier:2007wz} have been studied.

\par
In this paper, we make a precise calculation for the process \pphww
at the LHC including the contributions of the QCD NLO corrections
and the gluon-gluon fusion subprocess, for the purpose of avoiding a
possible experimentally observed deviation from the LO prediction
due to the QCD effects being misinterpreted. As we shall see from
the following investigation that these QCD NLO corrections and the
contribution from the gluon-gluon fusion process turn out to be
potentially important in observations of the signal of \pphww
process and should be taken into account in experimental data
analysis. In section II we give the calculation description of the
LO cross section of \pphww process, and the calculations of the
complete QCD NLO radiative contribution and the correction from
gluon-gluon fusion subprocess are provided in section III. In
section IV we present some numerical results and discussion, and
finally a short summary is given.

\par
\section{The LO cross section of the \pphww process}
\par
In the LO and higher order calculations we employ FeynArts3.4
package\cite{fey} to generate Feynman diagrams and their
corresponding amplitudes. The amplitude calculations are implemented
by applying FormCalc5.4 programs\cite{formloop}.

\par
\begin{figure*}
\begin{center}
\includegraphics*[125pt,420pt][530pt,710pt]{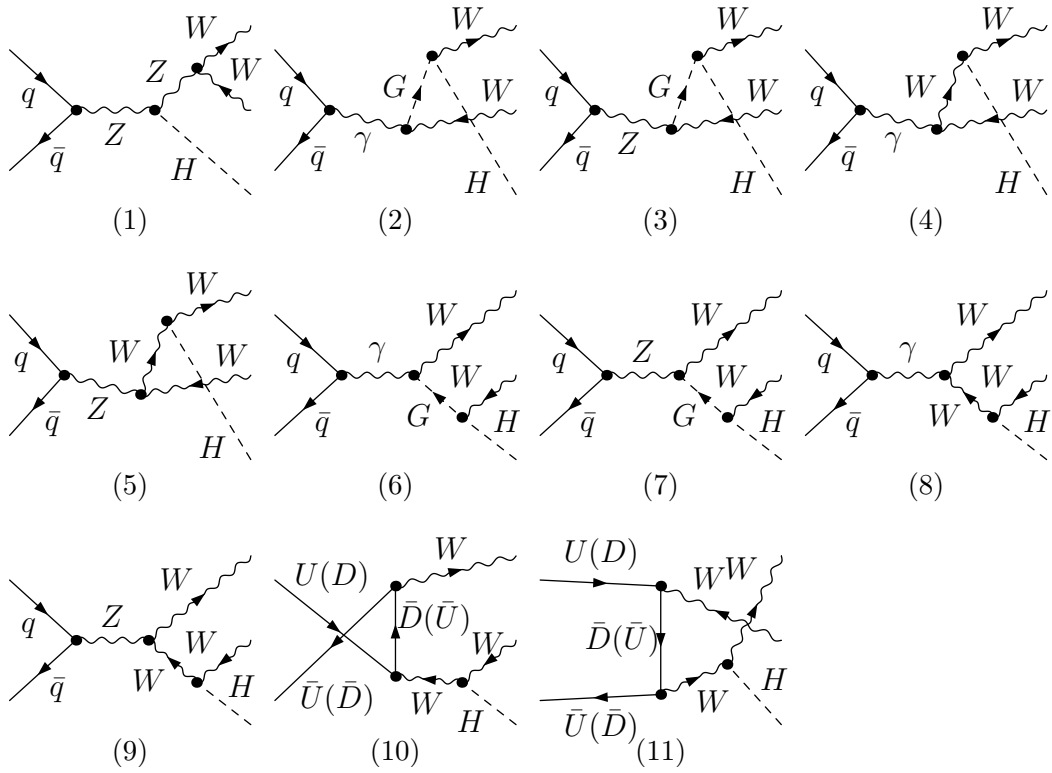}
\caption{\label{fig1} The tree-level Feynman diagrams for the
\qqhww($q=u,d,s,c$, $U=u,c$, $D=d,s$) subprocess, which are
considered in our LO calculations. }
\end{center}
\end{figure*}

The leading order contribution to the cross section of the parent
process \pphww comes from the subprocess of $H^0W^+W^-$ production
via quark-antiquark$(q=u,d,s,c)$ annihilation. We denote the
subprocess as
\begin{equation}
\label{process1} q(p_1)+\bar q(p_2) \to H^0(p_3)+
W^+(p_4)+W^-(p_5),~~~~(q=u,d,s,c).
\end{equation}
where $p_{1}$, $p_{2}$ and $p_{3}$, $p_{4}$, $p_{5}$ represent the
four-momenta of the incoming partons and the outgoing $H^0$,
$W^{\pm}$ bosons, respectively. We use the 't Hooft-Feynman gauge in
our LO calculations, if there is no other statement. We ignore the
contribution from the Feynman diagrams which involve the couplings
between fermions(u-, d-, s-, or c-quarks) and Higgs boson, since the
Yukawa coupling strength is proportional to fermion mass and the
masses of u-, d-, s-, and c-quark are relatively small and can be
negligible. The Feynman diagrams for the subprocess \qqhww at the LO
are depicted in Fig.\ref{fig1},

\par
The expression for the LO cross section for the subprocess \qqhww
has the form as
\begin{eqnarray}
\label{LO}\hat{\sigma}^{0}_{q\bar q}= \frac{1}{4}\frac{1}{9}\frac{(2
\pi )^4}{2\hat{s}}\int \sum_{spin}^{color} |{\cal M}_{LO}|^2
d\Omega_{3}
\end{eqnarray}
where the factors $\frac{1}{4}$ and $\frac{1}{9}$ come from the
averaging over the spins and colors of the initial partons
respectively, $\hat{s}$ is the partonic center-of-mass energy
squared, and ${\cal M}_{LO}$ is the amplitude of all the tree-level
diagrams shown in Fig.\ref{fig1}. The summation is taken over the
spins and colors of all the relevant particles in the \qqhww
subprocess. The integration is performed over the three-body phase
space of the final particles $H^0$, $W^{+}$ and $W^{-}$. The
phase-space element $d\Omega_{3}$ in Eq.(\ref{LO}) is expressed as
\begin{eqnarray}\label{PhaseSpace}
{d\Omega_{3}}=\delta^{(4)} \left( p_1+p_2-\sum_{i=3}^5 p_i \right)
\prod_{j=3}^5 \frac{d^3 \textbf{\textsl{p}}_j}{(2 \pi)^3 2 E_j}.
\end{eqnarray}

\par
Within the framework of the QCD factorization, the LO cross section
for the process \ppqqhww at the LHC can be obtained by performing
the following integration of the cross section for the subprocess
\qqhww over the partonic luminosities (see Eq.(\ref{integration})).
\begin{equation}
\label{integration} \sigma_{LO}= \sum_{ij=u\bar u}^{d\bar d,s\bar
s,c\bar c} \int_{0}^{1}dx_1 \int_{0}^{1} dx_2 \left[
G_{i/P_1}(x_1,\mu_f) G_{j/P_2}(x_2,\mu_f)+(x_1 \leftrightarrow x_2,
P_1 \leftrightarrow P_2)\right] \hat{\sigma}^{0}_{i j}(\hat{s}=x_1
x_2 s),
\end{equation}
where $G_{i/A}(x,\mu_f)$ is the parton($i=u,d,s,c$) distribution
function of proton $A(=P_1,P_2)$ which describes the probability to
find a parton $i$ with momentum $xp_A$ in proton $A$, $s$ is defined
as the total colliding energy squared in proton-proton collision,
$\hat{s}=x_1x_2 s$, and $\mu_f$ is the factorization energy scale.
In our LO calculations, we adopt the CTEQ6L1\cite{pdfs} parton
distribution functions.

\par
\section{QCD corrections }
\par
At the leading order, the parent process \pphww involves four
subprocesses, i.e., \qqhww, where $q=u,d,c$ and $s$. Due to the poor
luminosities for charm- and strange-quarks in protons, the
contribution to the LO cross section for the parent process \pphww
from the subprocesses $s\bar s, c\bar c \to H^0W^+W^-$ is relatively
small. Our calculation shows their contribution part to the LO cross
section is less than $10\%$ at the LHC. Therefore, in the
calculations beyond the LO we consider reasonably only the QCD
corrections to the processes $pp \to u\bar u, d\bar d \to
H^0W^+W^-+X$.

\par
Our QCD correction to the \pphww process at the LHC can be divided
into two parts: One is the QCD virtual correction, which should be
considered together with the contribution from the real
gluon/light-quark emission subprocesses in order to cancel the
soft/collinear IR singularities appeared in the virtual correction.
Actually, there still exists remaining collinear divergency which
can be absorbed by the parton distribution functions. Another part
is from the gluon-gluon fusion subprocess which gives the
contribution to the cross section of the \pphww process ${\cal
O}(\alpha_s)$ order higher than that of the previous subprocess at
the QCD NLO.

\par
\subsection{Virtual corrections to the subprocess \qqhww }
\par
\begin{figure*}
\begin{center}
\includegraphics*[125pt,510pt][530pt,710pt]{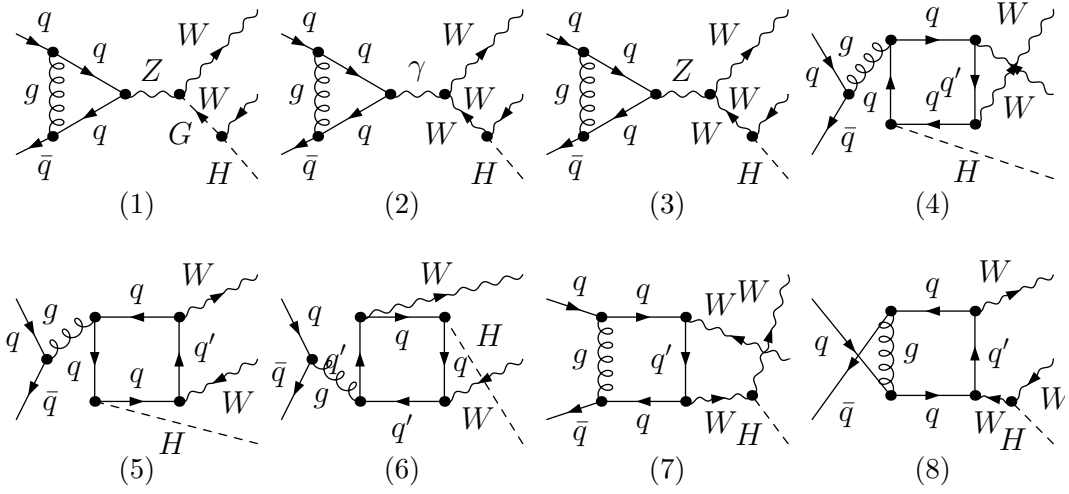}
\vspace*{-0.3cm} \centering \caption{\label{fig2} Some of the
one-loop Feynman diagrams for the subprocess \qqhww($q\bar q=u\bar
u,d\bar d$). }
\end{center}
\end{figure*}

In our calculations, all the divergences are regularized by using
the dimensional regularization method in $D=4-2 \epsilon$ dimensions
and the modified minimal subtraction ($\overline{\rm MS}$) scheme is
applied to renormalize the relevant fields. There are 171 virtual
QCD NLO diagrams contributing to the subprocess \qqhww in the SM,
including self-energy(94), vertex(35), box(5) and counterterm(37)
diagrams. We present part of these diagrams in Fig.\ref{fig2}. There
exist both ultraviolet(UV) and soft/collinear infrared(IR)
singularities in the calculation of the one-loop diagrams, but the
total QCD NLO amplitude of subprocess \qqhww is UV finite after
performing renormalization procedure. Nevertheless, it still
contains soft/collinear IR singularities as shown in
Eq.(\ref{virtual cross section}).

\begin{eqnarray}
\label{virtual cross section} d\hat{\sigma}^V_{q\bar
q}=d\hat{\sigma}^{0}_{q\bar q} \left[\frac{\alpha_s}{2 \pi}
\frac{\Gamma(1-\epsilon)}{\Gamma(1-2 \epsilon)}\left(\frac{4 \pi
\mu_r^2}{\hat{s}}\right)^{\epsilon}\right]
\left(\frac{A_{2}^{V}}{\epsilon^2}+\frac{A_{1}^{V}}{\epsilon}
+A_{0}^V\right),~~(q=u,d)
\end{eqnarray}
where
\begin{eqnarray}
A_{2}^{V}&=& -2 C_F,~~~~A_{1}^{V} = -3 C_F,~~~~C_F=4/3.
\end{eqnarray}

\par
As we shall see later that the soft/collinear IR singularities can
be cancelled by adding the contributions of the \qqhwwg and \qghwwq
subprocesses, and redefining the parton distribution functions at
the NLO. In the numerical calculations of the virtual corrections,
we use the expressions in Refs.\cite{OneTwoThree,Four,Five} to
implement the numerical evaluations of IR safe one-point, 2-point,
3-point, 4-point and 5-point integrals.

\par
\subsection{Real gluon emission subprocess \qqhwwg  }
\par
We denote the $q-\bar q(q=u,d)$ annihilation subprocess with a real
gluon emission as
\begin{equation}
q(p_1)+\bar q (p_2) \to H^0(p_3) +W^+(p_4) + W^-(p_5) + g(p_6).
\end{equation}

\par
\begin{figure*}
\begin{center}
\includegraphics*[125pt,410pt][530pt,715pt]{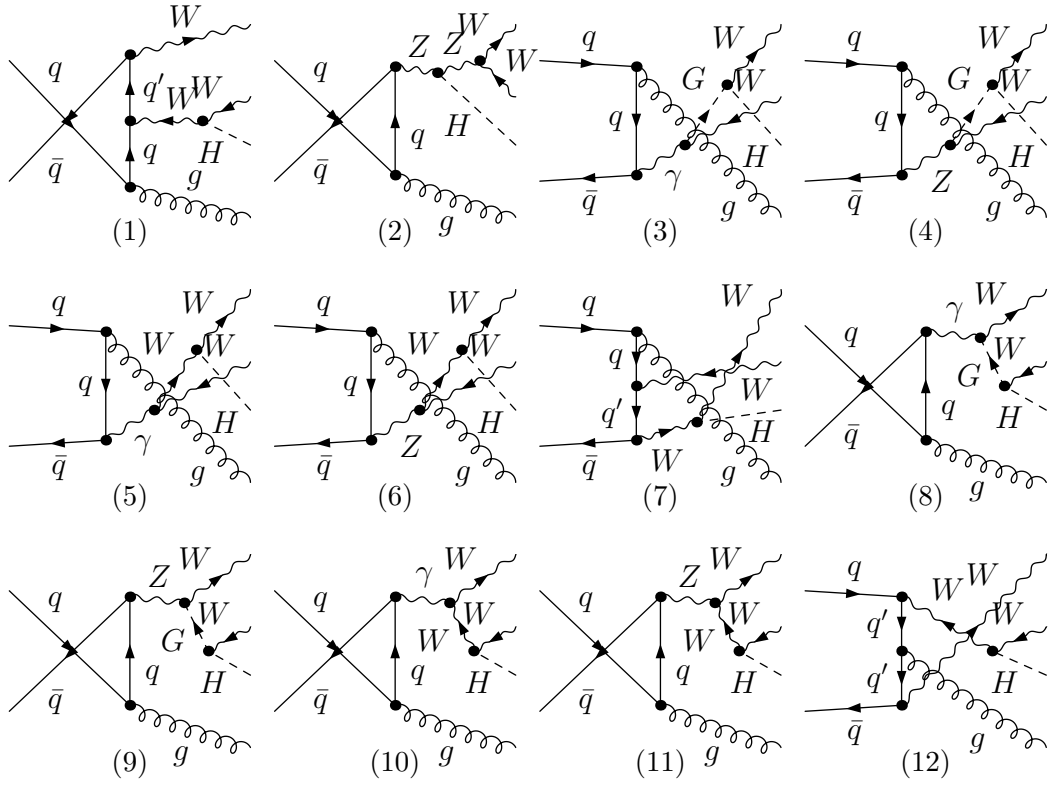}
\vspace*{-0.3cm} \centering \caption{\label{fig3} The tree-level
Feynman diagrams for the real gluon emission subprocess
\qqhwwg$(q=u,d)$. }
\end{center}
\end{figure*}

The real gluon emission subprocess \qqhwwg (shown in Fig.\ref{fig3})
produces both soft and collinear IR singularities which can be
conveniently isolated by adopting the two cutoff phase space slicing
(TCPSS) method\cite{TCPSS}. The soft IR singularity in the
subprocess \qqhwwg at the LO cancels the analogous singularity
arising from the one-loop level virtual corrections to the \qqhww
subprocess.

\par
In performing the calculations with the TCPSS method, we should
introduce arbitrary small soft cutoff $\delta_s$ and collinear
cutoff $\delta_c$. The phase space of the \qqhwwg subprocess can be
split into two regions, $E_6 \leq \delta_s\sqrt{\hat{s}}/2$(soft
gluon region) and $E_6
> \delta_s\sqrt{\hat{s}}/2$(hard gluon region) by soft cutoff $\delta_s$.
The hard gluon region is separated as hard collinear(${\rm HC}$) and
hard non-collinear ($\overline{\rm HC}$) regions by cutoff
$\delta_c$. The ${\rm HC}$ region is the phase space where
$-\hat{t}_{16}$(or $-\hat{t}_{26}$)$<\delta_c \hat{s}$
$(\hat{t}_{16}\equiv(p_1-p_6)^2$ and
$\hat{t}_{26}\equiv(p_2-p_6)^2)$. Therefore, the cross section for
this real gluon emission subprocess can be expressed as
\begin{eqnarray}
\hat{\sigma}^R_g(q\bar{q} \to
H^0W^+W^-g)=\hat{\sigma}^S_g+\hat{\sigma}^H_g &=&
\hat{\sigma}^S_g+\hat{\sigma}^{\rm HC}_g+\hat{\sigma}^{\overline{\rm
HC}}_g.
\end{eqnarray}
\par
The differential cross section for the subprocess \qqhwwg in the
soft region is given as
\begin{eqnarray}
\label{soft cross section} d\hat{\sigma}^S_g(q\bar q \to H^0W^+W^-g
)=d\hat{\sigma}^0_{q\bar q} \left[\frac{\alpha_s}{2 \pi}
\frac{\Gamma(1-\epsilon)}{\Gamma(1-2 \epsilon)}\left(\frac{4 \pi
\mu_r^2}{\hat{s}}\right)^{\epsilon}\right]
\left(\frac{A^S_2}{\epsilon^2}+\frac{A^S_1}{\epsilon}+A^S_0 \right),
\end{eqnarray}
with
\begin{eqnarray}
A^S_2&=& 2 C_F,~~~A^S_1 = - 4 C_F \ln \delta_s, ~~~A^S_0 = 4 C_F
\ln^2 \delta_s.
\end{eqnarray}
\par
The differential cross section for the process $pp \to q\bar q \to
H^0W^+W^-g+X$, $d\sigma^{HC}_g$ in the hard collinear region, can be
written as
\begin{eqnarray}
\label{collinear-g} d\sigma^{HC}_g&=&d\hat{\sigma}^0_{q\bar q}
\left[\frac{\alpha_s}{2 \pi} \frac{\Gamma\left (1-\epsilon\right
)}{\Gamma(1-2 \epsilon)}\left(\frac{4 \pi
\mu_r^2}{\hat{s}}\right)^{\epsilon}\right] \left
(-\frac{1}{\epsilon}\right )\delta_c^{-\epsilon} \left \{
P_{qq}(z,\epsilon)[G_{q/P_1}(x_1/z)G_{\bar{q}/P_2}(x_2) \right. \nb \\
&+& \left. G_{\bar{q}/P_1}(x_1/z)G_{q/P_2}(x_2)] +
(x_1\leftrightarrow x_2, P_1 \leftrightarrow P_2)\right \}
\frac{dz}{z}\left (\frac{1-z}{z}\right )^{-\epsilon}dx_1dx_2,
\end{eqnarray}
where $G_{q(\bar{q})/P}(x, \mu_f)$ is the bare parton distribution
function of quark(anti-quark) and P refers to proton.
$P_{qq}(z,\epsilon)$ is the $D$-dimensional unregulated ($z<1$)
splitting function which can be written explicitly as
\begin{eqnarray}
P_{qq}(z,\epsilon)=P_{qq}(z)+ \epsilon P'_{qq}(z),~~~ P_{qq}(z)=C_F
\frac{1+z^2}{1-z},~~~ P'_{qq}(z)=-C_F (1-z).
\end{eqnarray}

\vskip 15mm
\subsection{Real light-quark emission subprocess \qghwwq }
\par
\begin{figure*}
\begin{center}
\includegraphics*[120pt,260pt][530pt,665pt]{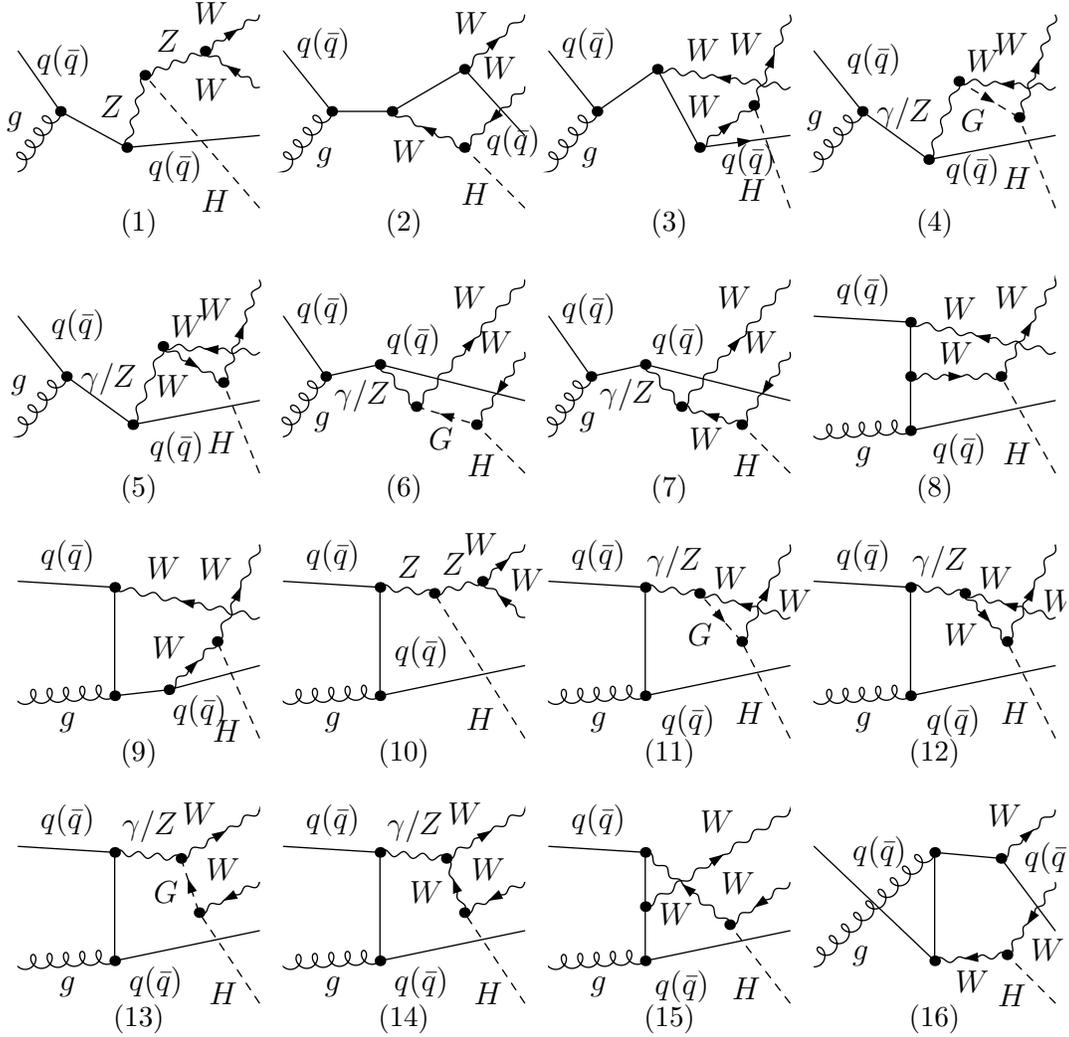}
\vspace*{-0.3cm} \centering \caption{\label{fig4} The tree-level
Feynman diagrams for the real light-quark emission subprocesses
\qghwwq$(q=u,d)$. }
\end{center}
\end{figure*}

Beside the real gluon emission subprocess discussed above, there is
another kind of contribution called the real light-quark emission
subprocess which has the same order contribution with previous real
gluon emission subprocess in perturbation theory. The corresponding
Feynman diagrams of the subprocesses \qghwwq$(q=u,d)$ at the
tree-level are shown in Fig.\ref{fig4}.

\par
These subprocesses contain only the initial state collinear
singularities. Using the TCPSS method described above, we split the
phase space into collinear region and non-collinear region by
introducing a cutoff $\delta_c$. Then the cross sections for the
subprocesses $qg \rightarrow H^0W^+W^-q$ and $\bar{q}g \rightarrow
H^0W^+W^-\bar{g}$ can be expressed as
\begin{equation}
\hat{\sigma}^R(qg \rightarrow H^0W^+W^-q) = \hat{\sigma}^R_q =
\hat{\sigma}^{HC}_q + \hat{\sigma}^{\overline{HC}}_q
\end{equation}
\begin{equation}
\hat{\sigma}^R(\bar{q}g \rightarrow H^0W^+W^-\bar{q}) =
\hat{\sigma}^R_{\bar{q}} = \hat{\sigma}^{HC}_{\bar{q}} +
\hat{\sigma}^{\overline{HC}}_{\bar{q}}
\end{equation}

The cross sections $\hat{\sigma}^{\overline{HC}}_{q}$ and
$\hat{\sigma}^{\overline{HC}}_{\bar{q}}$ in the non-collinear region
are finite and can be evaluated in four dimensions using Monte Carlo
method. The differential cross section in the collinear region for
the processes $pp \to qg \to H^0W^+W^-q + X$, $d\sigma^{HC}_q$, can
be expressed as
\begin{eqnarray}
\label{collinear-d} d\sigma^{HC}_q&=&d\hat{\sigma}^0_{q\bar q}
\left[\frac{\alpha_s}{2 \pi} \frac{\Gamma(1-\epsilon)}{\Gamma(1-2
\epsilon)}\left(\frac{4 \pi
\mu_r^2}{\hat{s}}\right)^{\epsilon}\right ] \left (
-\frac{1}{\epsilon}\right )\delta_c^{-\epsilon}
P_{qg}(z,\epsilon)\left[ G_{g/P_1}(x_1/z)G_{q/P_2}(x_2) \right . \nb \\
&+& \left . (x_1\leftrightarrow x_2, P_1 \leftrightarrow P_2) \right
] \frac{dz}{z}\left (\frac{1-z}{z}\right )^{-\epsilon}dx_1dx_2,
\end{eqnarray}
The expression of the $d\sigma^{HC}_{\bar q}$ for the $pp \to
\bar{q}g \to H^0W^+W^-\bar{q} +X$ process, can be obtained by doing
the replacement of $G_{q/P_2}(x_2) \to G_{{\bar q}/P_2}(x_2)$ in the
right-handed side of Eq.(\ref{collinear-d}). In above equation
$G_{q(\bar{q})/P}(x)$ is the bare parton distribution function of
quark(anti-quark) in proton and
\begin{eqnarray}
P_{qg}(z,\epsilon)=P_{qg}(z)+ \epsilon P'_{qg}(z), ~~~
P_{qg}(z)=\frac{1}{2}[z^2+(1-z)^2],~~~ P'_{qg}(z) = -z(1-z).
\end{eqnarray}

\par
\subsection{Gluon-gluon fusion subprocess \gghww }
\par
\begin{figure*}
\begin{center}
\includegraphics*[130pt,415pt][530pt,710pt]{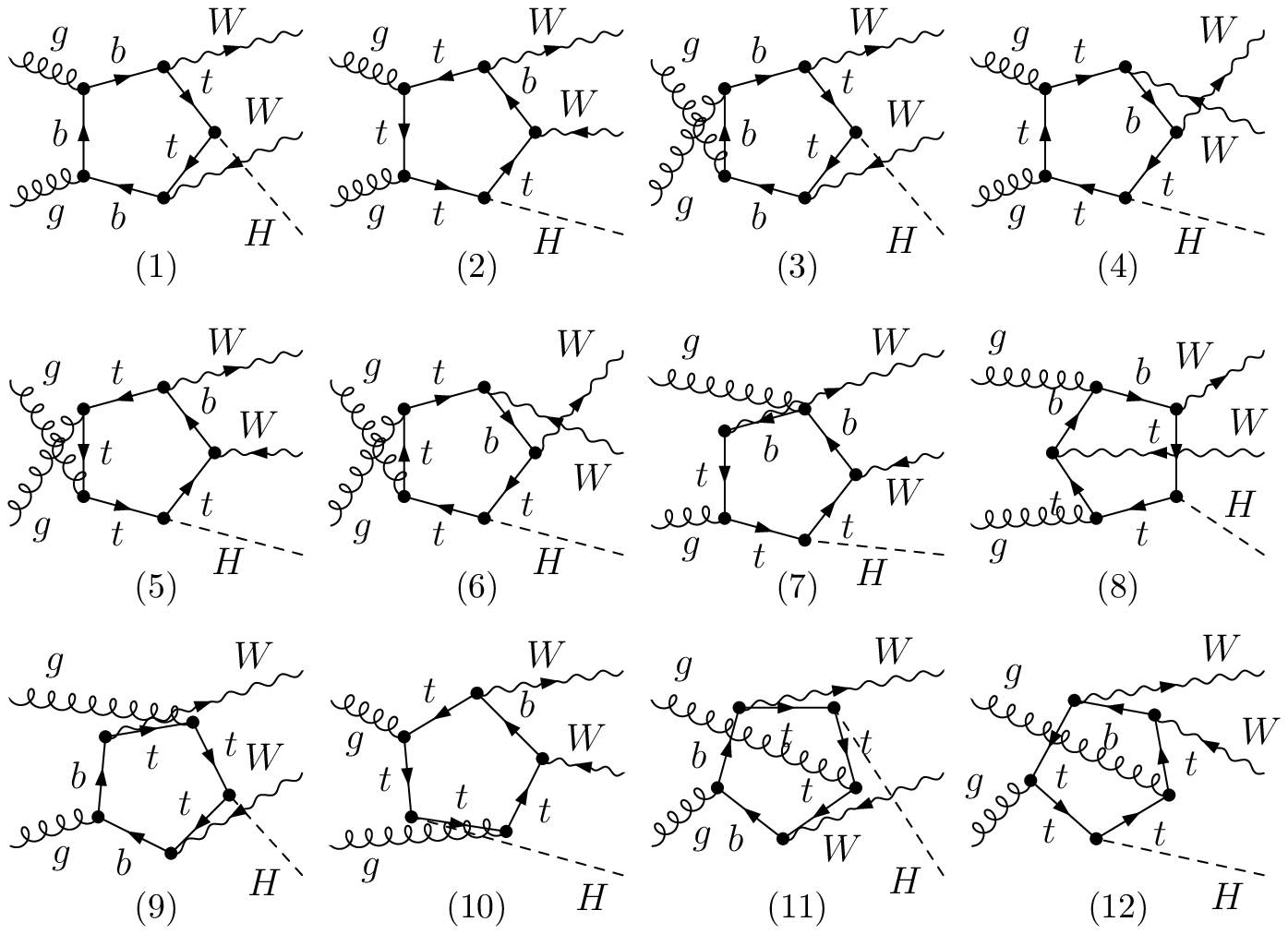}
\vspace*{-0.3cm} \centering \caption{\label{fig5} The pentagon
diagrams for the \gghww subprocess. }
\end{center}
\end{figure*}

The lowest order contribution of the \gghww subprocess is at the
one-loop level. This contribution to the process \pphww is ${\cal
O}(\alpha_s)$ order higher than the QCD NLO corrections from the
one-loop process \ppqqhww, the production rate of the $pp \to gg \to
H^0W^+W^-+X$ could be non-negligible, due to the large gluon
luminosity in TeV-scale proton-proton collision at the LHC. Here we
include the contribution of the gluon-gluon fusion subprocess in the
calculations of the QCD corrections to the \pphww process. We
neglect again the Feynman diagrams involving the interaction between
light fermions and Higgs boson. Among all the 292 QCD one-loop
Feynman diagrams, there are 63 self-energy, 148 vertex, 69 box and
12 pentagon diagrams. All the pentagon diagrams for the \gghww
subprocess are depicted in Fig.\ref{fig5} as a presentation.

\par
Again we employ the aforementioned dimensional regularization to
isolate the UV and IR divergences in one-loop calculation. Since
there is no tree-level diagram for the \gghww, the calculation for
this subprocess can be simply carried out by summing all
unrenormalized reducible and irreducible one-loop diagrams, and we
find the numerical results are UV and IR finite. We get the lowest
order differential cross section of the subprocess \gghww expressed
as:
\begin{eqnarray}
d\hat {\sigma}_{gg} &=& \frac{1}{4}\frac{1}{64} \frac{(2 \pi
)^4}{4|\vec{p}_1|\sqrt{s}} \sum_{spin}^{color} |{\cal M}_{gg}|^2
d\Omega_{3}.
\end{eqnarray}
where factors $1/4$ and $1/64$ are obtained by taking averages of
the initial spins and colors, and the phase space element of
three-body final states, $d\Omega_{3}$, is defined as in
Eq.(\ref{PhaseSpace}).

\par
After integration of $d\hat {\sigma}_{gg}$ over the partonic
luminosities, we can see from the numerical results that although
the contributions from the subprocess \gghww are much smaller than
the QCD NLO corrections to the \ppqqhww process, the QCD relative
correction from the \ppgghww at the LHC is significant and can even
reach $24\%$ when $m_H=160~GeV$.

\par
\subsection{QCD corrected cross section for the \pphww process }
\par
After adding the renormalized virtual corrections and the real
gluon/light-quark emission corrections to the subprocess \qqhww, the
partonic cross sections still contain the collinear divergences,
which can be absorbed into the redefinition of the distribution
functions at the NLO. Using the $\overline{\rm MS}$ scheme, the
scale dependent NLO parton distribution functions are given as
\begin{eqnarray}
G_{i/P}(x,\mu_f)&=&G_{i/P}(x)+\sum_{j=q,\bar q,g}\left
(-\frac{1}{\epsilon}\right )\left[\frac{\alpha_s}{2 \pi}
\frac{\Gamma(1-\epsilon)}{\Gamma(1-2 \epsilon)}\left(\frac{4 \pi
\mu_r^2}{\mu_f^2}\right)^{\epsilon}\right]\int^1_z\frac{dz}{z}P_{ij}(z)G_{j/P}(x/z),
\nb \\
&&~~~~~~~~~~~~~~~~~~~~~~~~~~~~~~~~~~~~~~~~~~~(q=u,d,~i=u,\bar
u,d,\bar d,g).
\end{eqnarray}
By using above definition, we get the QCD counter-terms of parton
distribution function which are combined with the hard collinear
contributions to result in the $O(\alpha_s)$ expression for the
remaining collinear contributions:
\begin{eqnarray}
\label{collinear cross section}
d\sigma^{coll}&=&\sum_{q=u,d}d\hat{\sigma}^0_{q\bar q}
\left[\frac{\alpha_s}{2 \pi} \frac{\Gamma(1-\epsilon)}{\Gamma(1-2
\epsilon)}\left(\frac{4 \pi
\mu_r^2}{\hat{s}}\right)^{\epsilon}\right]\left \{
\tilde{G}_{q/P_1}(x_1,\mu_f)G_{\bar{q}/P_2}(x_2,\mu_f) \right .\nb
\\&+& \left . \tilde{G}_{\bar{q}/P_1}(x_1,\mu_f)G_{q/P_2}(x_2,\mu_f) +
\left [\frac{A_1^{sc}(q \to q g)}{\epsilon}+A_0^{sc}(q \to q
g)\right ]\cdot \right .\nb \\
&&\left . \cdot G_{q/P_1}(x_1,\mu_f)G_{\bar{q}/P_2}(x_2,\mu_f) +
(x_1 \leftrightarrow x_2, P_1 \leftrightarrow P_2)\right \}dx_1dx_2,
\end{eqnarray}
where
\begin{eqnarray}
A_1^{sc}(q \to qg)&=& C_F(2 \ln \delta_s+3/2), ~~~A_0^{sc} =
A_1^{sc} \ln(\frac{\hat{s}}{\mu_f^2}),
\end{eqnarray}
and
\begin{eqnarray}
\tilde{G}_{q/P}(x,\mu_f)=\sum_{j=q,g}\int^{1-\delta_s \delta_{qj}}_x
\frac{dy}{y}G_{j/P}(x/y,\mu_f)\tilde{P}_{q j}(y),
\end{eqnarray}
with
\begin{eqnarray}
\tilde{P}_{ij}(y)=P_{ij} \ln\left
(\delta_c\frac{1-y}{y}\frac{\hat{s}}{\mu_f^2}\right )-P'_{ij}(y).
\end{eqnarray}
We can find that the sum of the soft (expressed in Eq.(\ref{soft
cross section})), collinear(expressed in Eq.(\ref{collinear cross
section})), and ultraviolet renormalized virtual correction
(expressed in Eq.(\ref{virtual cross section})) terms is finite,
i.e.,
\begin{eqnarray}
A^S_2+A^V_2=0,~~~~~ A^S_1+A^V_1+ 2 A_1^{sc}(q\to qg)=0.
\end{eqnarray}
The final result for the total QCD correction($\Delta\sigma^{QCD}$)
consists of a three-body term $\Delta\sigma^{(3)}$ and a four-body
term $\Delta\sigma^{(4)}$.
\begin{eqnarray} \label{3body}
\Delta\sigma^{(3)}&=&\frac{\alpha_s}{2 \pi} \sum_{q=u,d} \int
dx_1dx_2d\hat{\sigma}^0_{q\bar q}
\left \{ G_{q/P_1}(x_1,\mu_f)G_{\bar q/P_2}(x_2,\mu_f) [A^S_0+A^V_{0}+2 A_0^{sc}(q\to qg)] \right. \nb \\
&+& \left. \tilde{G}_{q/P_1}(x_1,\mu_f)G_{\bar
q/P_2}(x_2,\mu_f)+G_{q/P_1}(x_1,\mu_f)\tilde{G}_{\bar
q/P_2}(x_2,\mu_f)+(x_1 \leftrightarrow x_2,
P_1 \leftrightarrow P_2 )\right \}\nb \\
&+&  \frac{1}{2}\int dx_1dx_2d\hat{\sigma}_{gg}\left \{
G_{g/P_1}(x_1,\mu_f)G_{g/P_2}(x_2,\mu_f)+(x_1 \leftrightarrow x_2,
P_1 \leftrightarrow P_2)\right\}.
\end{eqnarray}
And
\begin{eqnarray}\label{4body}
\Delta\sigma^{(4)}&=&\sum_{q=u,d}\int dx_1dx_2
[G_{q/P_1}(x_1,\mu_f)G_{\bar{q}/P_2}(x_2,\mu_f)+(x_1 \leftrightarrow
x_2, P_1 \leftrightarrow P_2)]\hat{\sigma}_{g}^{\overline{\rm
HC}}(\hat{s}=x_1 x_2 s) \nb \\
&+& \sum_{q=u,d}^{\bar {u},\bar {d}}\int dx_1dx_2
[G_{q/P_1}(x_1,\mu_f)G_{g/P_2}(x_2,\mu_f)+(x_1 \leftrightarrow x_2,
P_1 \leftrightarrow P_2)]\hat{\sigma}_{q}^{\overline{\rm
HC}}(\hat{s}=x_1 x_2 s). \nb \\
&&~~~~~~~~~
\end{eqnarray}
where $\hat{\sigma}^{\overline{\rm HC}}_g(\hat{s}=x_1 x_2 s)$ is the
cross section for the subprocess \qqhwwg($q=u,d$) in the hard
non-collinear phase space region at the colliding energy
$\hat{s}=x_1 x_2 s$ in the partonic center-of-mass system.
$\hat{\sigma}^{\overline{\rm HC}}_q(\hat{s})$, where $q=u,d,\bar
u,\bar d$, represent the cross sections in the non-collinear phase
space regions for the subprocesses $u g \to H^0W^+W^-u$, $d g \to
H^0W^+W^-d$, $\bar u g \to H^0W^+W^-\bar u$ and $\bar d g \to
H^0W^+W^-\bar d$, respectively.

\par
Finally, the QCD corrected total cross section for the \pphww
process is
\begin{eqnarray}
\sigma^{QCD}=\sigma^{0}+\Delta\sigma^{QCD}=\sigma^{0}+\Delta\sigma^{(3)}+\Delta\sigma^{(4)}.
\end{eqnarray}
where the LO cross section part of the parent process \pphww is
expressed as
\begin{eqnarray}
\label{sigma0} \sigma^{0}=\sum_{q=u,d}^{s,c}\int dx_1 dx_2
d\hat{\sigma}^0_{q\bar q} \left \{
G_{q/P_1}(x_1,\mu_f)G_{\bar{q}/P_2}(x_2,\mu_f)+(x_1 \leftrightarrow
x_2, P_1 \leftrightarrow P_2)\right \}.
\end{eqnarray}

\par
In our numerical calculations by using Eq.(\ref{sigma0}) for the $pp
\to u\bar u,d \bar d,gg \to H^0W^+W^-+X$ processes, we use the
CTEQ6M\cite{pdfs} parton distribution functions, while for the $pp
\to s\bar s,c \bar c \to H^0W^+W^-+X$ processes, we adopt the
CTEQ6L1 distribution functions.

\par
\section{Numerical results and discussion}
\par
In this section we describe and discuss the numerical results of our
calculations for the \pphww process at the LO, the corrections at
the QCD NLO and the contribution from the gluon-gluon fusion
subprocess. We take one-loop and two-loop running $\alpha_{s}(\mu)$
for the LO and the higher order calculations,
respectively\cite{hdata}. We set the factorization scale and the
renormalization scale being equal, and take $\mu\equiv\mu_f = \mu_r
= (m_H+2 m_W)/2$ by default unless otherwise stated, the CKM matrix
being a unit matrix. We adopt $m_u=m_d=m_g=0$ and employ the
following numerical values for the relevant input parameters:
\cite{hdata}
\begin{equation}
\begin{array}{lll}  \label{input1}
\alpha(m_Z)^{-1}=127.918, &m_W=80.398~{\rm GeV},    &m_Z=91.1876~{\rm GeV}, \\
m_t=171.2~{\rm GeV},      &m_s=104~{\rm MeV}, &m_c=1.27~{\rm GeV}, \\
m_b=4.2~{\rm GeV}. &  &
\end{array}
\end{equation}

\par
By taking $m_H=120~GeV$ and the CTEQ6L1 parton distribution
functions, we perform a check for the correctness of the LO
calculation of the process \ppuuhww. We use the
FeynArts3.4/FormCalc5.4\cite{fey,formloop} packages and
CompHEP-4.4p3 program\cite{CompHEP}, and apply the Feynman and
unitary gauges, separately. The numerical results are listed in
Table \ref{tab1}. We can see that all those results are in good
agreement.
\begin{table}
\begin{center}
\begin{tabular}{|c|c|c|c|}
\hline  $\sigma_{LO}$(fb) & $\sigma_{LO}$(fb) & $\sigma_{LO}$(fb)& $\sigma_{LO}$(fb)   \\
\hline    CompHEP &  CompHEP & FeynArts  & FeynArts \\
\hline  Feynman Gauge & unitary gauge & Feynman Gauge & unitary gauge \\
\hline  5.902(4) & 5.903(4) & 5.898(6)& 5.898(6) \\
\hline
\end{tabular}
\end{center}
\begin{center}
\begin{minipage}{15cm}
\caption{\label{tab1} The numerical results of the LO cross sections
for the process \ppuuhww by using FeynArts3.4/FormCalc5.4 packages
and CompHEP-4.4p3 program, adopting the Feynman and unitary gauges
separately, with the CTEQ6L1 parton distribution functions and
$m_H=120~GeV$. }
\end{minipage}
\end{center}
\end{table}

\par
\begin{figure}[htbp]
\includegraphics[width=3.2in,height=3in]{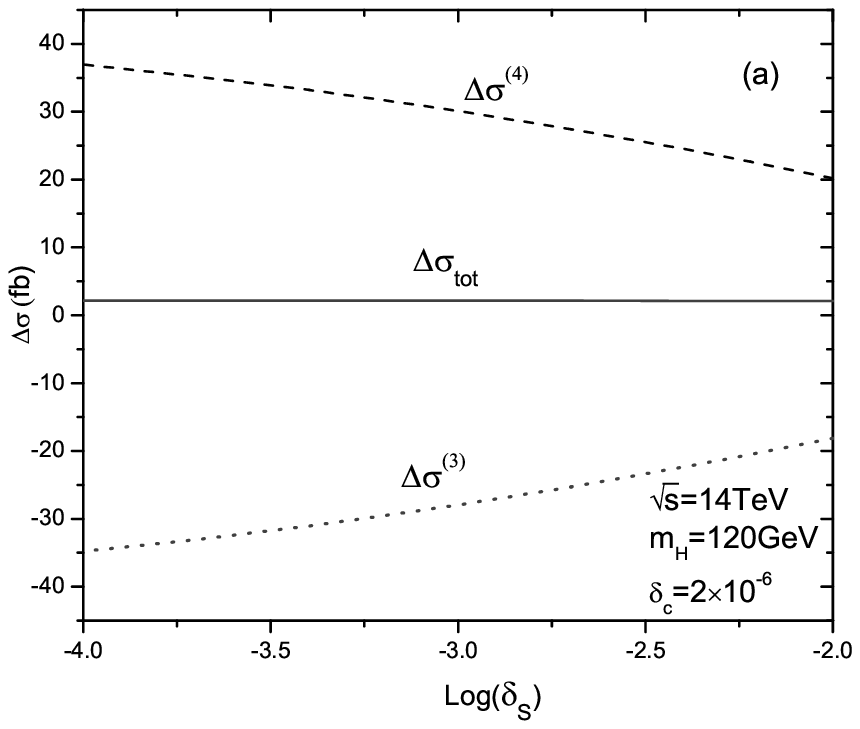}%
\hspace{0in}%
\includegraphics[width=3.2in,height=3in]{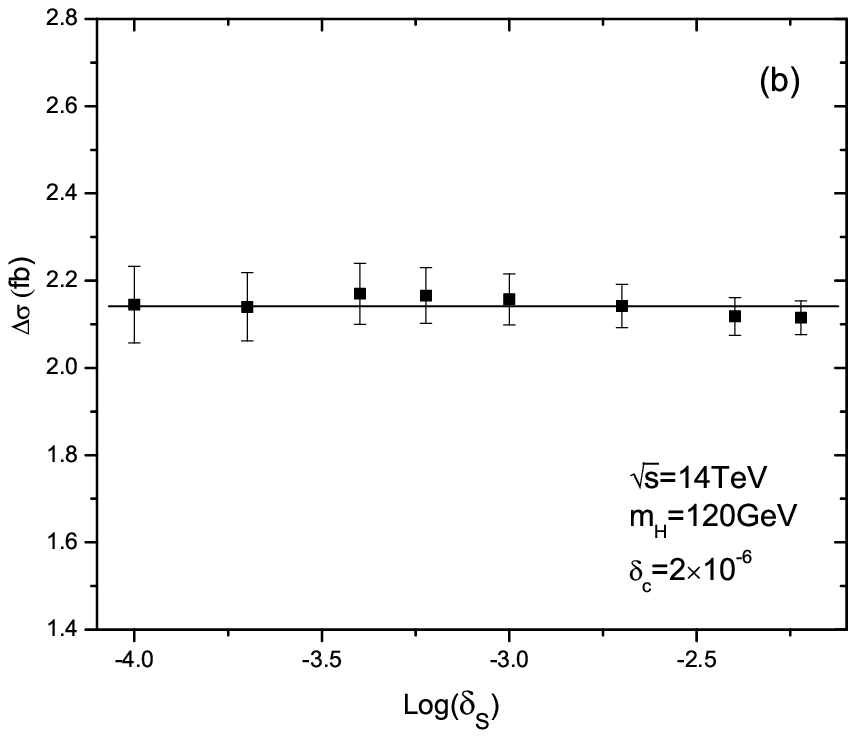}%
\hspace{0in}%
\caption{\label{fig6} (a) The dependence of QCD NLO correction parts
to the \ppuuhww process on the soft cutoff $\delta_s$ at the LHC
with $m_H=120~GeV$, the collinear cutoff $\delta_c = 2 \times
10^{-6}$ and $\sqrt{s}=14~TeV$. (b) The amplified curve for the
total QCD correction $\Delta\sigma^{QCD}$ to the process \ppuuhww,
where it includes the calculation errors. }
\end{figure}
\begin{figure}[htbp]
\includegraphics[width=3.2in,height=3in]{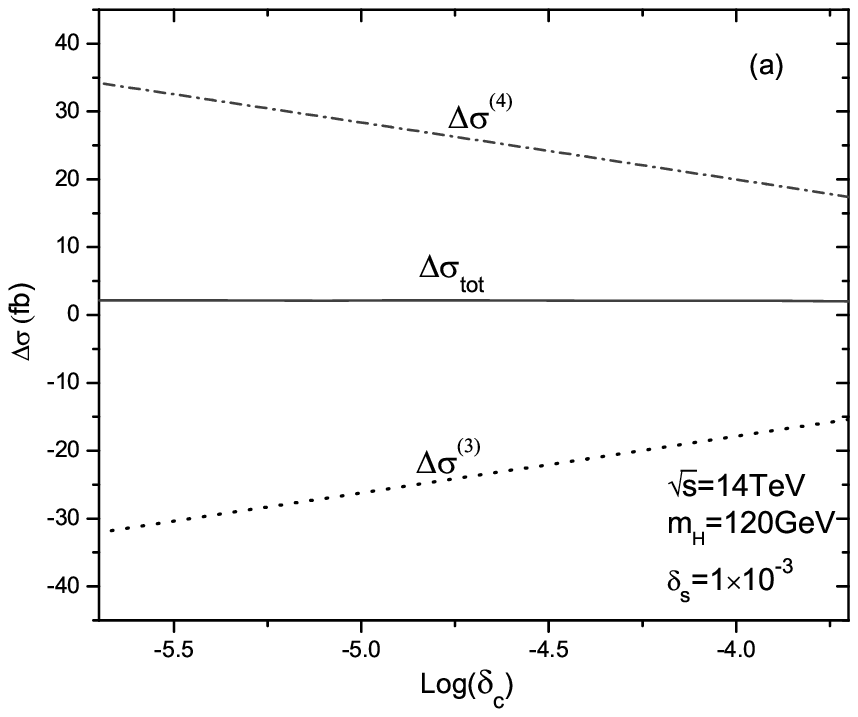}%
\hspace{0in}%
\includegraphics[width=3.2in,height=3in]{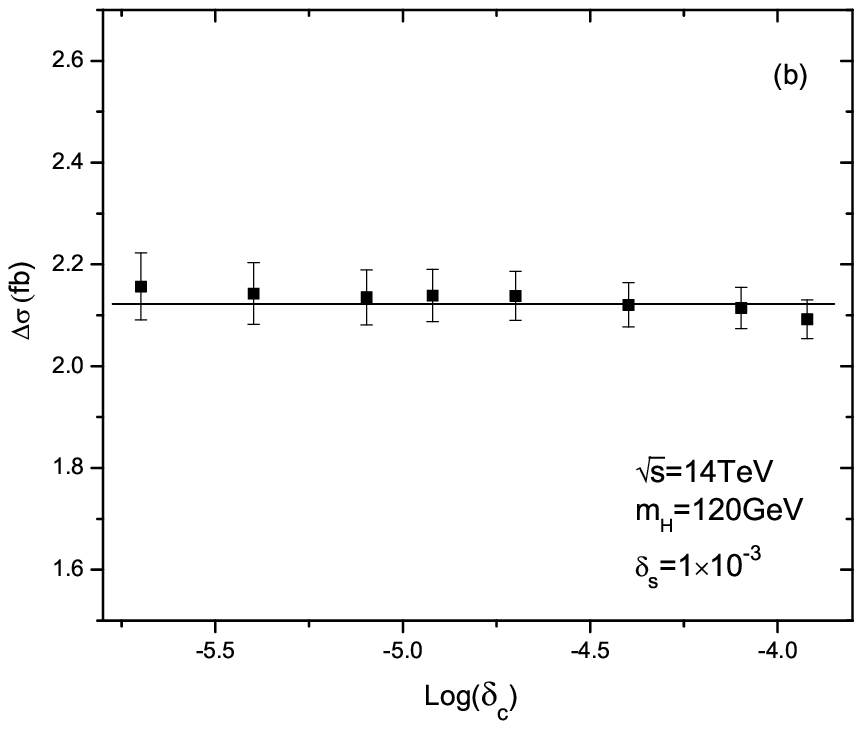}%
\hspace{0in}%
\caption{\label{fig7} (a) The dependence of the QCD NLO correction
parts to the \ppuuhww process on the collinear cutoff $\delta_c$ at
the LHC with $m_H=120~GeV$, $\delta_s=10^{-3}$ and
$\sqrt{s}=14~TeV$. (b) The amplified curve for the total QCD
correction $\Delta\sigma^{QCD}$ to the process \ppuuhww, where it
includes the calculation errors.}
\end{figure}

Figs.\ref{fig6}(a,b) show that our total QCD correction to the
\ppuuhww process does not depend on the arbitrarily chosen value of
the cutoff $\delta_s$ with the fixed value of $\delta_c = 2 \times
10^{-6}$ by adopting the TCPSS method. The three-body
correction($\Delta\sigma^{(3)}$, see Eq.(\ref{3body})) and four-body
correction($\Delta\sigma^{(4)}$, see Eq.(\ref{4body})) and the total
QCD correction ($\Delta\sigma^{QCD}$) for the \ppuuhww process at
the LHC, are depicted as the functions of the soft cutoffs
$\delta_s$ by taking $m_H=120~GeV$ and $\delta_s$ running from
$10^{-4}$ to $10^{-2}$ in Fig.\ref{fig6}(a). The amplified curve for
$\Delta\sigma^{QCD}$ is presented in Fig.\ref{fig6}(b) together with
calculation errors. While Figs.\ref{fig7}(a,b) show the independence
of the total QCD correction to the \ppuuhww process on the cutoff
$\delta_c$ where we take $\delta_s=10^{-3}$. In Fig.\ref{fig7}(b)
the amplified curve for $\Delta\sigma^{QCD}$ of the \ppuuhww process
is depicted. The fact that the total QCD correction
$\Delta\sigma^{QCD}$ for the \ppuuhww process is independent of
these two cutoffs, not only proofs the cancelation of soft/collinear
IR divergency in the total QCD correction for the process \ppuuhww,
but also partially verifies the correctness of our calculation. In
the following numerical calculations, we fix $\delta_s=10^{-3}$ and
$\delta_c=\delta_s/50$.

\par
\begin{figure}
\centering
\includegraphics[scale=0.75]{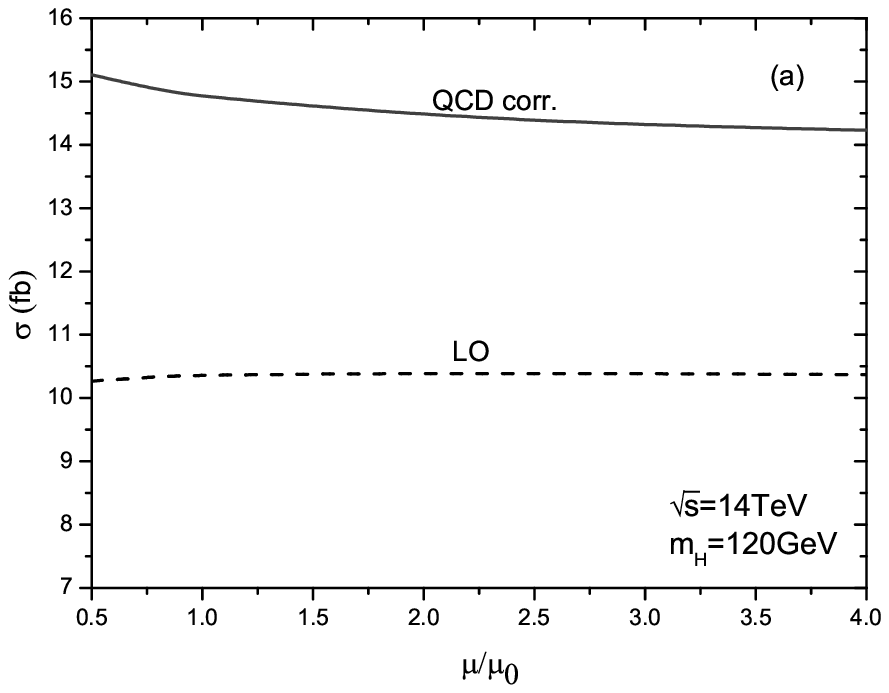}
\includegraphics[scale=0.75]{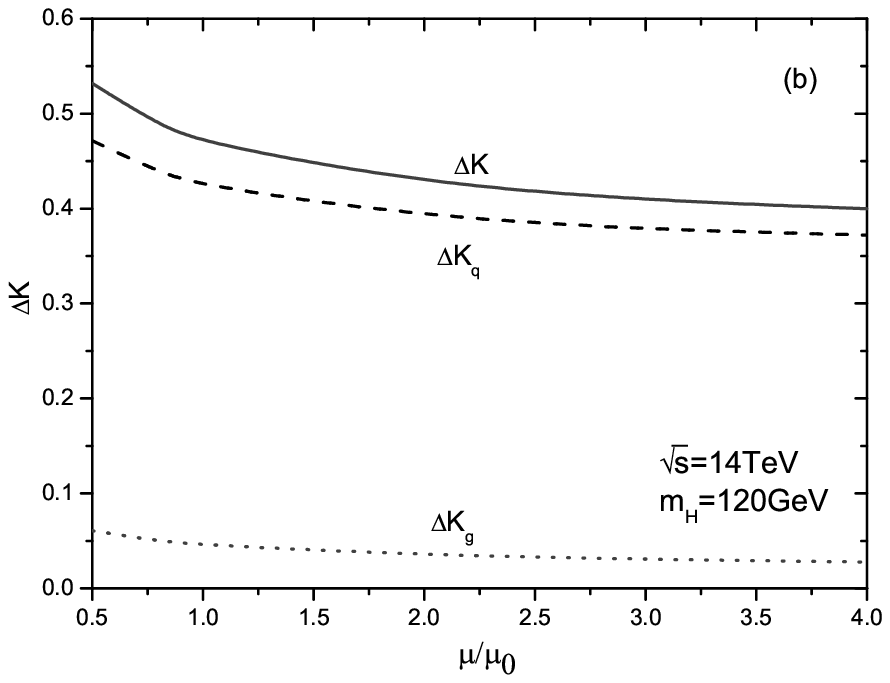}
\caption{\label{fig8} (a)The dependence of the LO and the QCD
corrected cross-sections for the process \pphww on the
factorization/renormalization scale($\mu/\mu_0$). (b)The total QCD
relative correction to the process \pphww($\Delta
K\equiv\Delta\sigma^{QCD}/\sigma_{LO}$), the QCD relative correction
parts from the \ppqqhww process($\Delta
K_{q}\equiv\Delta\sigma_q^{QCD}/\sigma_{LO}$) and the \ppgghww
process ($\Delta K_{g}\equiv\sigma_g^{QCD}/\sigma_{LO}$) versus the
factorization/renormalization scale($\mu/\mu_0$). Here we assume
$\mu \equiv \mu_f=\mu_r$ and define $\mu_0=(m_H+2 m_W)/2$. }
\end{figure}

In Figs.\ref{fig8}(a,b) we assumed $\mu \equiv \mu_r=\mu_f$ and
defined $\mu_0=(m_H+2 m_W)/2$. Fig.\ref{fig8}(a) shows the
dependence of the LO and the total QCD corrected cross-sections for
the process \pphww on the factorization/renormalization
scale($\mu/\mu_0$). We can see that the curve for LO cross section
has a tiny variation being less than one percent, but the variation
of the QCD corrected cross section is relative large by
approximately $10\%$ when the energy scale $\mu$ runs from
$0.5\mu_0$ to $4\mu_0$. It demonstrates that the LO curve
drastically underestimates the energy scale dependence of the QCD
correction. That is because there is no strong interaction in the LO
diagrams of the \qqhww subprocess, and its weak energy scale
dependence is the consequence of the parton distribution functions
being related to the factorization scale($\mu_f$). The similar
behavior is demonstrated in the Z production at the
Tevatron\cite{Anastasiou} and the production of three Z-bosons at
the LHC\cite{Achilleas}. Fig.\ref{fig8}(b) describes the total QCD
relative correction to the process \pphww defined as $\Delta
K\equiv\Delta\sigma^{QCD}/\sigma_{LO}$, the QCD relative corrections
from the NLO \ppqqhww and the LO \ppgghww processes, defined as
$\Delta K_{q}\equiv\Delta\sigma_q^{QCD}/\sigma_{LO}$ and $\Delta
K_{g}\equiv\sigma_g^{QCD}/\sigma_{LO}$ respectively, as the
functions of the factorization/renormalization scale($\mu/\mu_0$).
It demonstrates that the energy scale $\mu$ dependence of the cross
section for the \pphww process is mainly related to the
contributions of the QCD corrections to the \ppqqhww process, and
the dependence of the $\Delta K_{g}$ on the energy scale $\mu$ is
obviously weaker than the $\Delta K_{q}$.

\par
\begin{figure}
\centering
\includegraphics[scale=0.7]{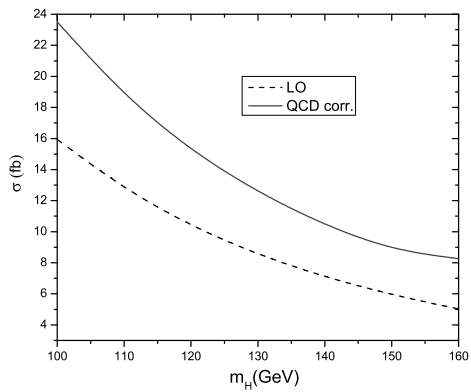}
\caption{\label{fig9}  The LO and the QCD corrected cross sections
for the process \pphww as the functions of the Higgs-boson mass
$m_H$ at the LHC. }
\end{figure}

In Fig.\ref{fig9} we present the plot of the LO and the QCD
corrected(including \ppgghww contribution) cross sections for the
process \pphww as the functions of the Higgs boson mass $m_H$ at the
LHC. From the figure we can see the cross sections at the LO and
including the QCD corrections are all sensitive to the Higgs boson
mass. We find the LO cross section decreases from $15.93~fb$ to
$5.03~fb$ and the QCD corrected cross section decreases from
$23.50~fb$ to $8.27~fb$ when $m_H$ goes up from $100~GeV$ to
$160~GeV$. And the corresponding
K-factor($K\equiv\sigma^{QCD}/\sigma_{LO}$) varies in the range from
$1.48$ to $1.64$.

\par
In Table \ref{tab2} we list some of the numerical results used in
Fig.\ref{fig9}. They are the data for the tree-level, the QCD
corrected(including the \ppgghww contribution) cross sections, the
total K-factor($K\equiv\frac{\sigma^{QCD}}{\sigma_{LO}}$) of the
process \pphww, the K-factor part contributed by the \ppqqhww
process up to ${\cal O}(\alpha^3\alpha_s)$ order($
K_{q}\equiv\frac{\sigma_q^{QCD}}{\sigma_{LO}}$) and the K-factor
part contributed by the \ppgghww process at the ${\cal
O}(\alpha^3\alpha_s^2)$ order($\Delta
K_{g}\equiv\frac{\sigma_g^{QCD}}{\sigma_{LO}}$) with the Higgs-boson
mass value being in the range from $100~GeV$ to $160~GeV$ at the
LHC. From Table \ref{tab2} we can see the LO and the QCD corrected
cross sections are all sensitive to the Higgs-boson mass, but the
total K-factor is not sensitive to the Higgs-boson mass except in
the vicinity where $m_H$ approaches to $2 m_W \sim 160~GeV$. The
contribution from the \ppgghww process to the total QCD corrections
can be remarkable at the LHC, and the QCD relative correction from
the process \ppgghww is generally about $4\%$, and can reach the
value of $24\%$ in the vicinity of $m_H \sim 160~GeV$, which is
about $37\%$ of the total QCD corrections. That large correction
enhancement at the position around $160~GeV$, is mainly induced by
the resonance effect of $m_H \sim 2 m_W$ occurring in those Feynman
diagrams for the subprocess \gghww, which involves a internal
Higgs-boson line interacting with two external W-bosons.
\begin{table}
\begin{center}
\begin{tabular}{|c|c|c|c|c|c|}
\hline $m_H (GeV)$ & $\sigma_{LO}(fb)$ & $\sigma^{QCD}(fb)$ &
$K_{q}$ & $\Delta K_{g}$  & $K $ \\
\hline 100 & 15.93(1) & 23.50(9)  & 1.435  & 0.040  &1.475 \\
\hline 110 & 12.763(8)& 18.75(7)  & 1.427  & 0.042  &1.469 \\
\hline 120 & 10.366(7)& 15.23(6)  & 1.424  & 0.045  &1.469 \\
\hline 130 & 8.522(6) & 12.53(5)  & 1.420  & 0.051  &1.471 \\
\hline 140 & 7.082(5) & 10.42(4)  & 1.413  & 0.059  &1.472 \\
\hline 150 & 5.941(4) & 8.83(3)   & 1.408  & 0.078  &1.486 \\
\hline 160 & 5.028(3) & 8.27(3)   & 1.403  & 0.241  &1.644 \\
\hline
\end{tabular}
\end{center}
\begin{center}
\begin{minipage}{15cm}
\caption{\label{tab2} The LO and the QCD corrected cross sections
for the \pphww process, the total
K-factor($K\equiv\frac{\sigma^{QCD}}{\sigma_{LO}}$) for the process
\pphww, the K-factor part contributed by the \ppqqhww process up to
${\cal O}(\alpha^3\alpha_s)$
order($K_q\equiv\frac{\sigma_q^{QCD}}{\sigma_{LO}}$) and the
K-factor part contributed by the \ppgghww process at the ${\cal
O}(\alpha^3\alpha_s^2)$ order($\Delta
K_g\equiv\frac{\sigma_g^{QCD}}{\sigma_{LO}}$) with the Higgs boson
mass value varying from $100~GeV$ to $160~GeV$ at the LHC. }
\end{minipage}
\end{center}
\end{table}

\par
\begin{figure}
\centering
\includegraphics[scale=0.75]{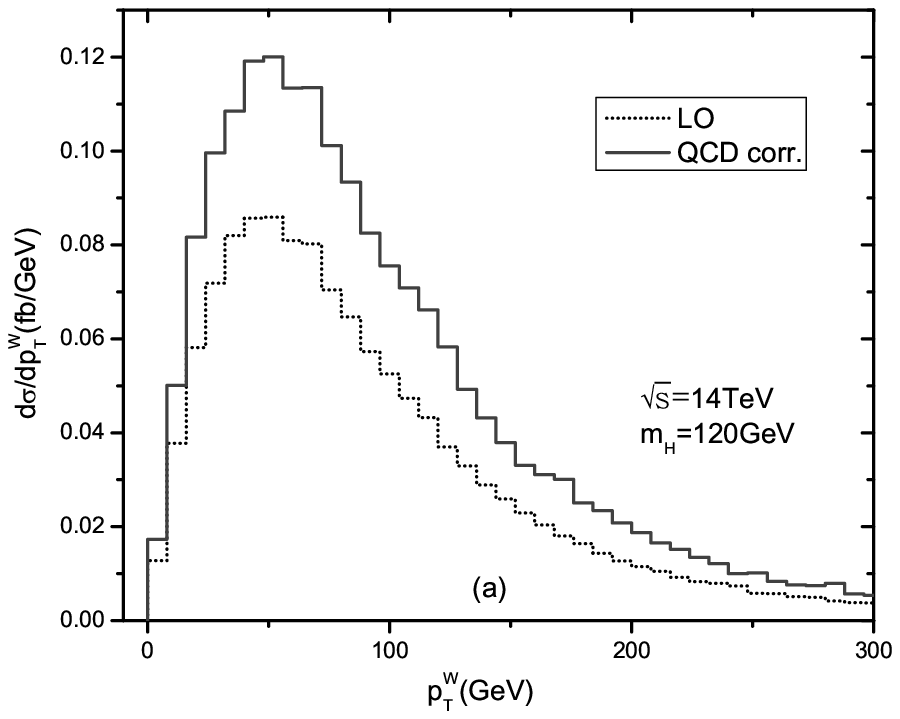}
\includegraphics[scale=0.75]{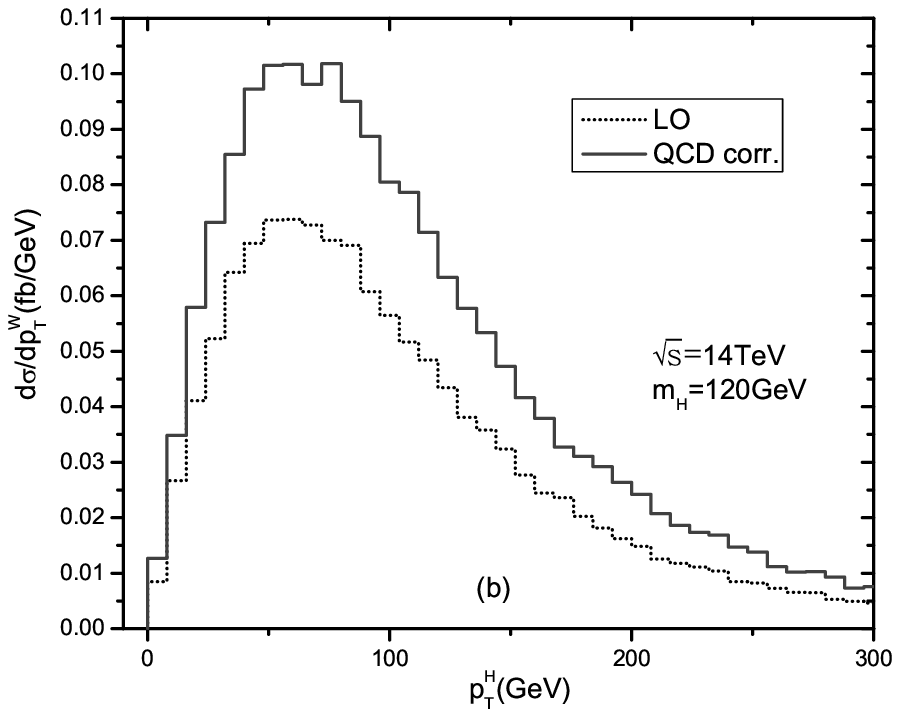}
\caption{\label{fig10} The distributions of the transverse momenta
of $W^+$- and $H^0$-boson for the \pphww process at the LO and
including QCD corrections at the LHC, by taking $m_H=120~GeV$. (a)
for the $W^+$-boson, (b) for the $H^0$-boson. }
\end{figure}

Since the distribution of the transverse momenta of $W^-$ boson is
the same as that of $W^+$ in the CP-conserving SM, we show only the
results for the transverse momentum distribution of $W^+$-boson
here. The differential cross sections of the $p_T$ for $W^+$-boson
at the LO and including the QCD corrections (QCD NLO correction to
the \pphww and the contribution of the \ppgghww process), i.e.,
$d\sigma_{LO}/dp_T^{W^+}$ and $d\sigma_{QCD}/dp_T^{W^+}$, are
depicted in Fig.\ref{fig10}(a), and the distributions of
$d\sigma_{LO}/dp_T^{H^0}$ and $d\sigma_{QCD}/dp_T^{H^0}$ for
$H^0$-boson are plotted in Fig.\ref{fig10}(b) separately, by taking
$m_H=120~GeV$. In both figures there exist peaks for the curves of
$p_T^{W^+}$ and $p_T^{H^0}$ at the LO and including QCD corrections.
All the peaks are located at the position around $p_T\sim 50~GeV$.
And we can see from Figs.\ref{fig10}(a-b) that both the differential
cross sections at the LO for $W^+$- and $H^0$-boson
($d\sigma_{LO}/dp_T^{W^+}$, $d\sigma_{LO}/dp_T^{H^0}$)), are
significantly enhanced by the QCD corrections.

\par
\begin{figure}
\centering
\includegraphics[scale=0.8]{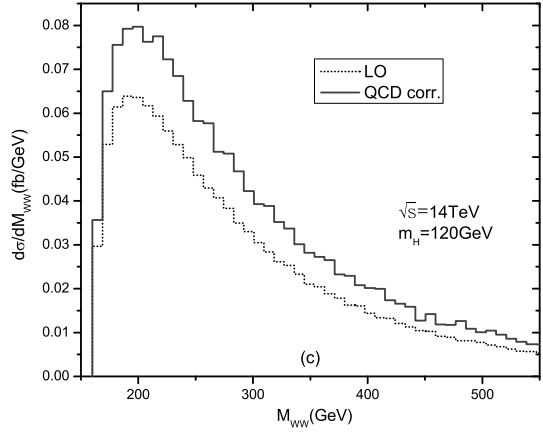}
\caption{\label{fig11} The distributions of the invariant mass of
W-pair at the LO and including QCD corrections at the LHC, when
$m_H=120~GeV$. }
\end{figure}

The curves for the distributions of W-pair invariant mass, denoted
as $M_{WW}$, at the LO and including the QCD corrections(involving
\ppgghww contribution), are drawn in Fig.\ref{fig11} respectively,
by taking $m_H=120~GeV$. The two curves show clearly that the QCD
correction including the QCD NLO correction part and contribution
from gluon-gluon fusion subprocess, enhances the LO differential
cross section $d\sigma_{LO}/dM_{WW}$ obviously in the plotted range
of $M_{WW}$, and the differential cross sections reach their maximal
values around the vicinity of $M_{WW}\sim 200~GeV$.

\vskip 10mm
\section{Summary}
\par
In this paper we investigate the phenomenological effects due to the
QCD NLO corrections and the gluon-gluon fusion subprocess in the
Higgs-boson production associated with a W-boson pair at the LHC. We
study the dependence of the LO and the QCD corrected cross sections
on the fctorization/renormalization energy scale and Higgs boson
mass. We present the LO and the QCD corrected distributions of the
transverse momenta of final particles and the differential cross
section of the $W$-pair invariant mass. We find that the QCD NLO
radiative corrections and the contribution from the \gghww
subprocess obviously modify the LO distributions, and the scale
dependence of the QCD corrected cross section is badly
underestimated by the LO results. Our numerical results show that
the K-factor of the QCD correction varies from $1.48$ to $1.64$ when
$m_H$ goes up from $100~GeV$ to $160~GeV$. We find also the cross
section of the \pphww process receives a remarkable QCD correction
from the contribution of \gghww subprocess at the LHC, and we should
consider this correction part in precise experimental data analyse.

\vskip 10mm
\par
\noindent{\large\bf Acknowledgments:} This work was supported in
part by the National Natural Science Foundation of
China(No.10875112, No.10675110).

\end{document}